\documentclass[titlepage,11pt]{article}
\usepackage{geometry} 
\usepackage{graphicx}
\newcommand{\uas}{$\mu$as}
\newcommand{\cgs}{erg~cm$^{-2}$~s$^{-1}$}

\begin{document}

\begin{titlepage}
       \vspace*{0.5cm}
        \noindent\makebox{\textbf{\LARGE The high energy universe at ultra-high resolution:}}
        \raisebox{-3mm}[0pt][0pt]{%
        \noindent\makebox{\textbf{\LARGE the power and promise of X-ray interferometry}}}
       
       \vspace{1.8cm}

        \noindent {\Large A White Paper submitted in response to ESA's Voyage 2050 call}
        
        \vspace{0.5cm}
        \large{For author list, see back page.}
        
        \vspace{0.6cm}
        \hspace{-0.5cm}
        \makebox{\includegraphics[width=0.43\textwidth]{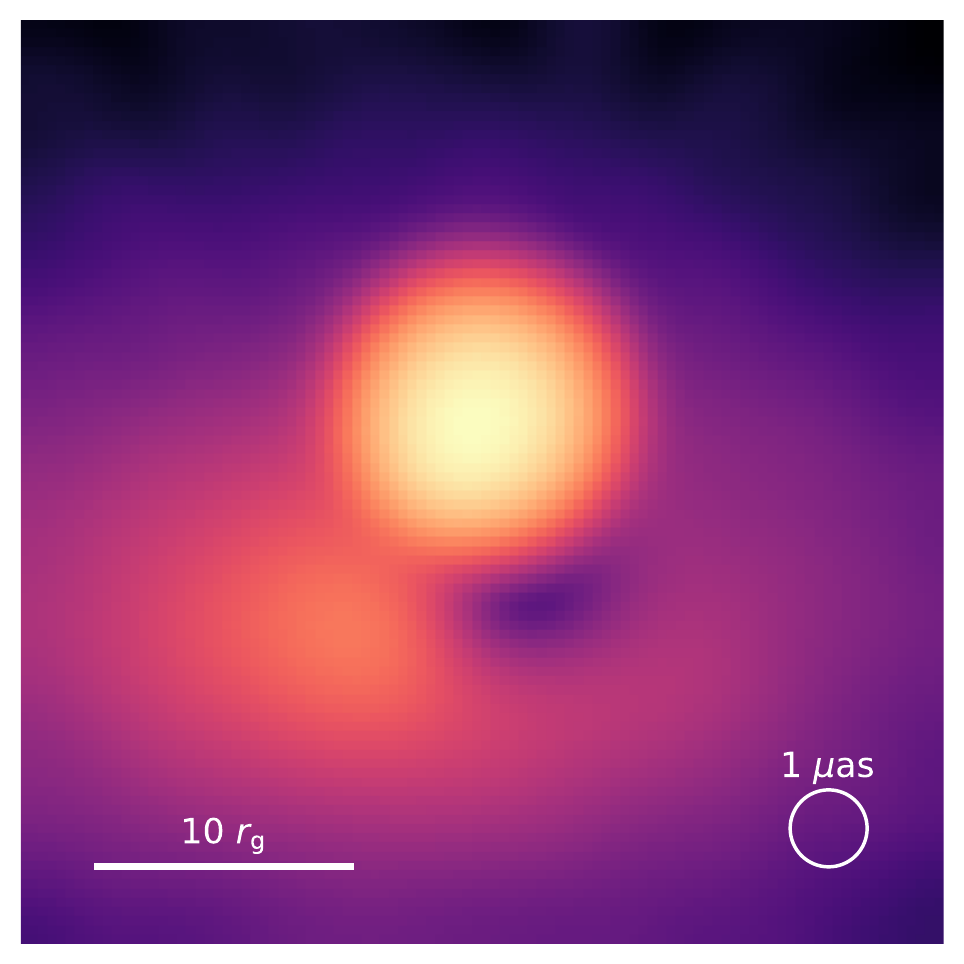}}
        \hspace{0.5cm}
        \raisebox{2mm}[0pt][0pt]{%
        \makebox{\includegraphics[width=0.41\textwidth]{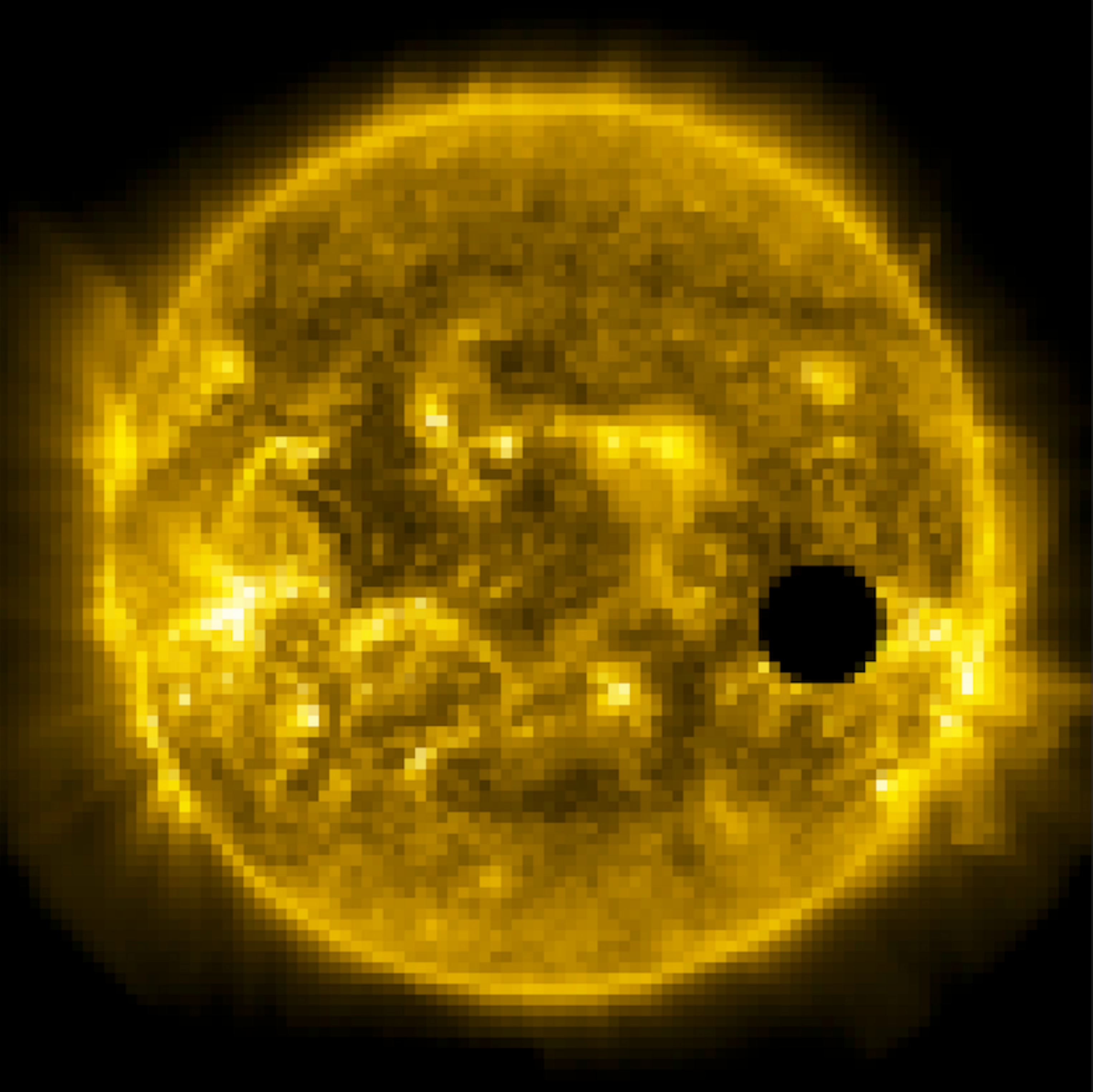}}}
        \vspace{0.5cm}
        
        \noindent \textbf{\Large Abstract} \\
        \noindent We propose the development of X-ray interferometry (XRI), to reveal the universe at high energies with ultra-high spatial resolution. With baselines which can be accommodated on a single spacecraft, XRI can reach 100 \uas\ resolution at 10~\AA\ (1.2 keV) and 20 \uas\ at 2~\AA\ (6 keV), enabling imaging and imaging-spectroscopy of (for example) X-ray coronae of nearby accreting supermassive black holes (SMBH) and the SMBH `shadow'; SMBH accretion flows and outflows; X-ray binary winds and orbits; stellar coronae within $\sim 100$~pc and many exoplanets which transit across them. For sufficiently luminous sources XRI will resolve sub-pc scales across the entire observable universe, revealing accreting binary SMBHs and enabling trigonometric measurements of the Hubble constant with X-ray light echoes from quasars or explosive transients. A multi-spacecraft `constellation' interferometer would resolve well below 1~\uas, enabling SMBH event horizons to be resolved in many active galaxies and the detailed study of the effects of strong field gravity on the dynamics and emission from accreting gas close to the black hole.
        
\end{titlepage}

\section{Context and power of X-ray interferometry}
Images have more power to inspire and change the direction of people's lives than any other tool at an astronomer's disposal.  Their public impact was spectacularly demonstrated in 2019 by the Event Horizon Telescope (EHT) image of the inner accretion flow and supermassive black hole ``shadow'' in M87, which made the front pages of the vast majority of global news media, reaching perhaps 4.5 billion people.  

The scientific impact of radically improving imaging capability is similarly enormous. In the sub-mm band, EHT's $\sim25$~microarcsecond (\uas) image of M87 revealed the first evidence for the central depression and ring of light associated with the accretion flow and photon orbits close to the black hole (BH) event horizon \cite{EHTm87L1_2019}, enabling stringent tests of the BH nature of the supermassive compact object \cite{EHTm87L5_2019}, a direct estimate of its mass \cite{EHTm87L6_2019} and a comparison with general relativistic magnetohydrodynamic (GRMHD) simulations of the accretion flow and jet, which implies that the jet is powered by extraction of rotational energy from the spinning BH \cite{EHTm87L5_2019}. Also recently, the GRAVITY instrument installed on the Very Large Telescope Interferometer (VLTI) has produced resolutions down to 2~mas in NIR and even more accurate astrometric measurements, enabling orbital motion of plasma blobs close to the event horizon of the SMBH in our Galactic centre to be detected \cite{GravitySgrorb2018}. Pushing to ultra-high resolutions allows the most extreme regimes of gravity and dynamics to be explored through images and dramatically reveals the details of seemingly well-studied astronomical objects.

The Rayleigh diffraction limit, $\theta_{R} = 1.22 \lambda/d$, provides the fundamental restriction on the angular resolution $\theta_{R}$ of a telescope of diameter $d$, which images light of wavelength $\lambda$.  The mirrors of optical/IR telescopes are already diffraction-limited from space, and these telescopes can approach the limit from the ground with the help of adaptive optics to cope with atmospheric disturbance.  Hence in the next decade, ESO's Extremely Large Telescope (ELT) will reach $\theta_{R} \sim 5$~mas with its 39~m mirror diameter. Significantly higher resolutions can be reached using interferometry, combining the signals from multiple telescopes to effectively circumvent the diffraction limit for a single telescope and reach resolutions $\theta_{I}= \lambda/2D$, where $D$ is the baseline between the telescopes. This approach is best known from radio astronomy where the simultaneous signals from different telescopes can be separately recorded and combined to enable sub-arcsec resolution with interferometric arrays such as JVLA and ALMA and sub-mas resolution with global Very Long Baseline Interferometry (VLBI). The latter's extension to the much shorter sub-mm wavelengths has led to the spectacular EHT result.

As work on optical/IR and sub-mm interferometry continues apace, an unexplored frontier remains in the X-ray band, which could yield resolutions comparable to those obtained by EHT with only a 1~m baseline interferometer, which combines X-ray beams using a combination of grazing-incidence flat mirrors\footnote{Nested, curved grazing-incidence mirrors which bring the X-rays to a focus are still very far from diffraction-limited, with significant improvements being in increased mirror size and not resolution.}, that would fit on board a single space-craft \cite{Willingale2004}. A `constellation' of multiple formation-flying spacecraft providing baselines of 100~m or more, would yield sub-\uas\ resolution \cite{Gendreauetal2009}. 

Microarcsecond imaging of astrophysical objects at X-ray wavelengths is uniquely powerful, for several reasons. Firstly, X-rays offer a unique probe of matter at a wide variety of temperatures and densities, including non-thermal plasmas, thermal plasma emission and atomic emission and absorption signatures due to X-ray reprocessing and absorption in colder gas. This opens up a much wider range of source-types and astrophysical processes to \uas\ X-ray imaging than is possible to study in other wavebands.  Furthermore, unlike detectors at longer wavelengths, X-ray imaging detectors can record individual photons over a broad energy range (e.g. 0.2-10~keV) with spectral resolution $R=\lambda/\Delta \lambda$ ranging from 10 (for CCDs at $\sim 1$~keV) to $>2500$ (at 6.4 keV for calorimeters such as the X-IFU on board {\it Athena}). Thus, \uas\ X-ray imaging could yield detailed broadband and high-spectral resolution spectra \emph{per image resolution element}, providing enormous diagnostic power on temperatures, composition and emission mechanisms, as well as dynamical information from line redshifts.

The compact nature of the brightest and most commonly-studied X-ray sources also makes them well-suited for ultra-high resolution \uas\ imaging, since they can be both resolved and their structure studied in detail at these scales. Accretion disks and the surrounding gas in nearby AGN subtend 100 $\mu$as scales, while pushing to sub-\uas\ can resolve the lensed event horizons of the SMBH in these AGN.  X-ray binary system orbits can be resolved at $\sim$kpc distances, enabling reprocessed X-ray emission from the outer disk, wind and companion star to be cleanly separated from the primary X-ray emission from the compact object, so that extended emission from outflows can be studied in detail and key parameters such as distance and mass measured with exquisite precision. At the other end of the luminosity scale, solar-type stars subtend $\sim 1$~mas at 10~pc, so that diffraction-limited X-ray images can easily resolve the coronal structures in nearby stars and can even resolve transiting exoplanets against the coronal background. 

There is substantial and continuing progress and development in X-ray detector technology (for high spectral and time resolution), as well as high-throughput but modest angular resolution focusing optics, but X-ray astronomy is significantly falling behind other wavebands in its potential for ultra-high resolution imaging.  Such a divergence could significantly limit future scientific advances, in an era when multi-wavelength measurements are increasingly important. Therefore, the time is right to begin serious planning and development towards realising the goal of \uas\ and sub-\uas\ resolution X-ray imaging via interferometry.  Work can already begin towards a single-spacecraft design that would enable 10--100~\uas\ imaging within the Voyage 2050 time-frame.  Our ambition, however, should be to also start the process of development to enable sub-\uas\ imaging and unlock the full potential of X-ray interferometry (XRI) using free-flying spacecraft.  As with many breakthroughs, there are serious hurdles that must first be overcome, not least in moving from a single spacecraft to a multi-spacecraft constellation, which is perhaps beyond the window of Voyage 2050 and even the career horizons of today's early career researchers.  However, the lesson from the remarkable recent rise of gravitational wave astronomy is that foundations need to be laid early on: putting off development towards a now-distant goal, will only push the achievement of that goal still further into the future.

In the remainder of this White Paper, we present the important and inspirational science that can be achieved by pushing ultra-high resolution X-ray imaging to what is achievable with a single-spacecraft interferometer, towards multiple spacecraft and ultimately a constellation.  We will end with a discussion of the technical feasibility of these different levels of ambition in X-ray interferomety, and a plausible roadmap towards their realisation.

\section{The promise of ultra-high resolution X-ray imaging}
\begin{figure}
    \centering
    \includegraphics[scale=0.45]{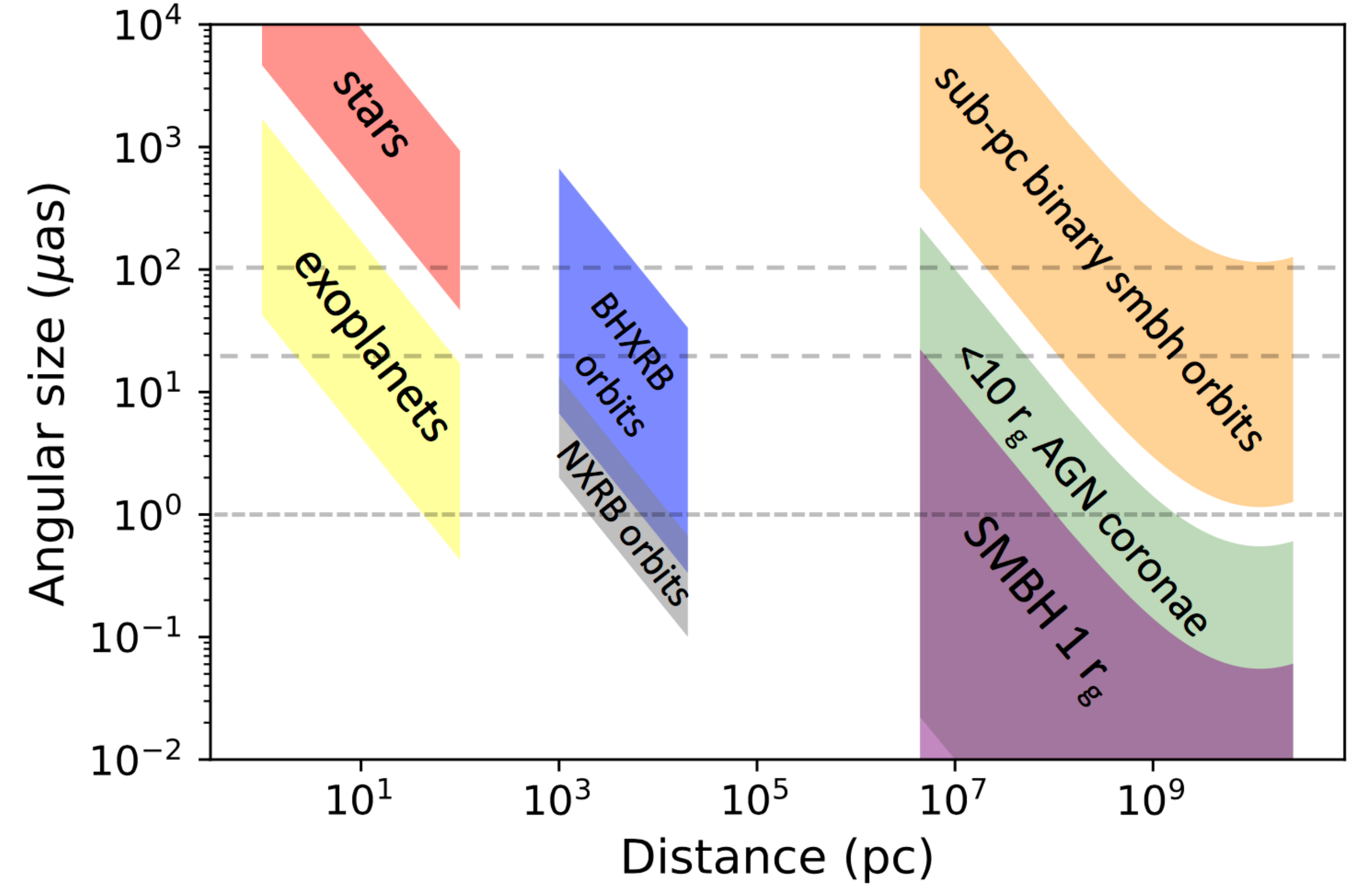}
    \caption{Summary plot showing typical source angular size range vs. distance, covering much of the X-ray interferometry science described in this paper. The upper and lower dashed lines show the resolution expected for a single-spacecraft XRI and 10\AA (1.2 keV) and 2\AA (6 keV) respectively.  The lower dotted line shows a notional 1\uas\ limit for effective BH horizon imaging, which a constellation XRI will easily surpass.}
    \label{fig:summaryplot}
\end{figure}

X-ray interferometers with resolutions from 100~\uas\ down to sub-\uas\ can address a wide range of fundamental physical and astrophysical questions, ranging from powerful explorations of General Relativity, accurate measurement of fundamental parameters, to the imaging of accretion inflows and outflows, stellar winds and coronae and even exoplanets (see Fig.~\ref{fig:summaryplot} for a summary). In this section we detail some of the unique science that will be enabled by ultra-high resolution X-ray imaging, which we address here in order of system energetics and/or physical scale rather than any preference or perceived value, since an X-ray interferometer of the kind we envisage would be a true observatory class instrument, rather than an experiment.

\subsection{Relativistic dynamics and tests of General Relativity}
Einstein's theory of General Relativity was proposed over a century ago and is still the standard framework for the description of gravitational fields and the geometry of the spacetime. While its predictions have been extensively tested with Solar System experiments and observations of radio pulsars \cite{Will2014}, the strong field regime is only beginning to be explored and there are many theoretical models that have the same predictions as General Relativity in the weak field limit and present deviations only when gravity becomes strong \cite{Berti2015}. Astrophysical BHs are ideal laboratories for exploring General Relativity in the strong field regime.

General Relativity with BHs can be explored either with electromagnetic \cite{Bambi2017} or gravitational wave techniques \cite{Yunesetal2016}. The two methods are complementary because they examine different aspects of the theory. Gravitational wave studies can better test the theory in settings with dynamic space-time, while electromagnetic methods are more suitable to test the interactions between matter and the gravitational field, when the space-time is static. For example, deviations from geodesic motion or variation of fundamental constants can leave a signature in the electromagnetic spectrum without affecting the gravitational wave one.

\begin{figure}[tb]
    \centering
    \includegraphics[width=0.53\textwidth]{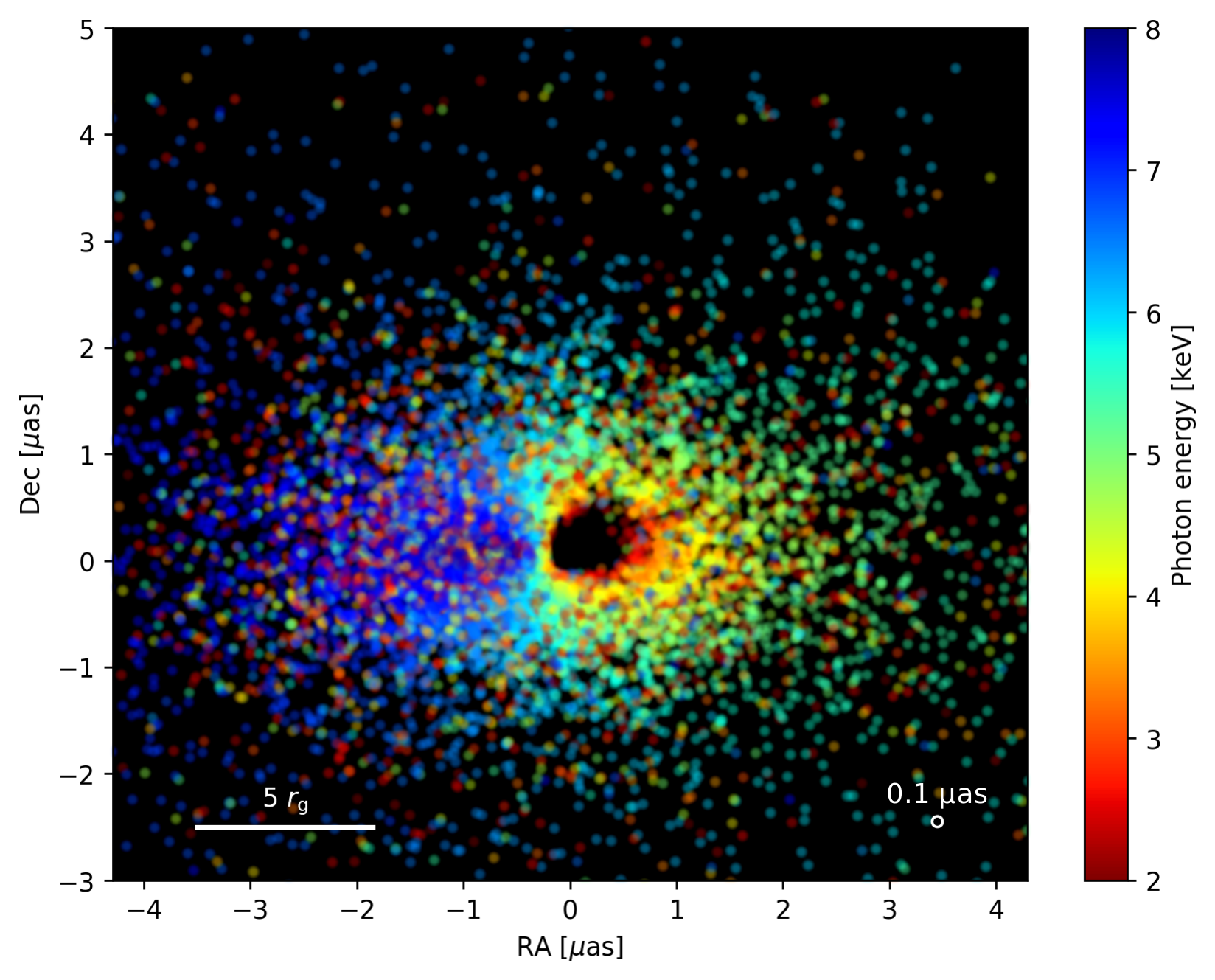}
    \hfill
    \includegraphics[width=0.43\textwidth]{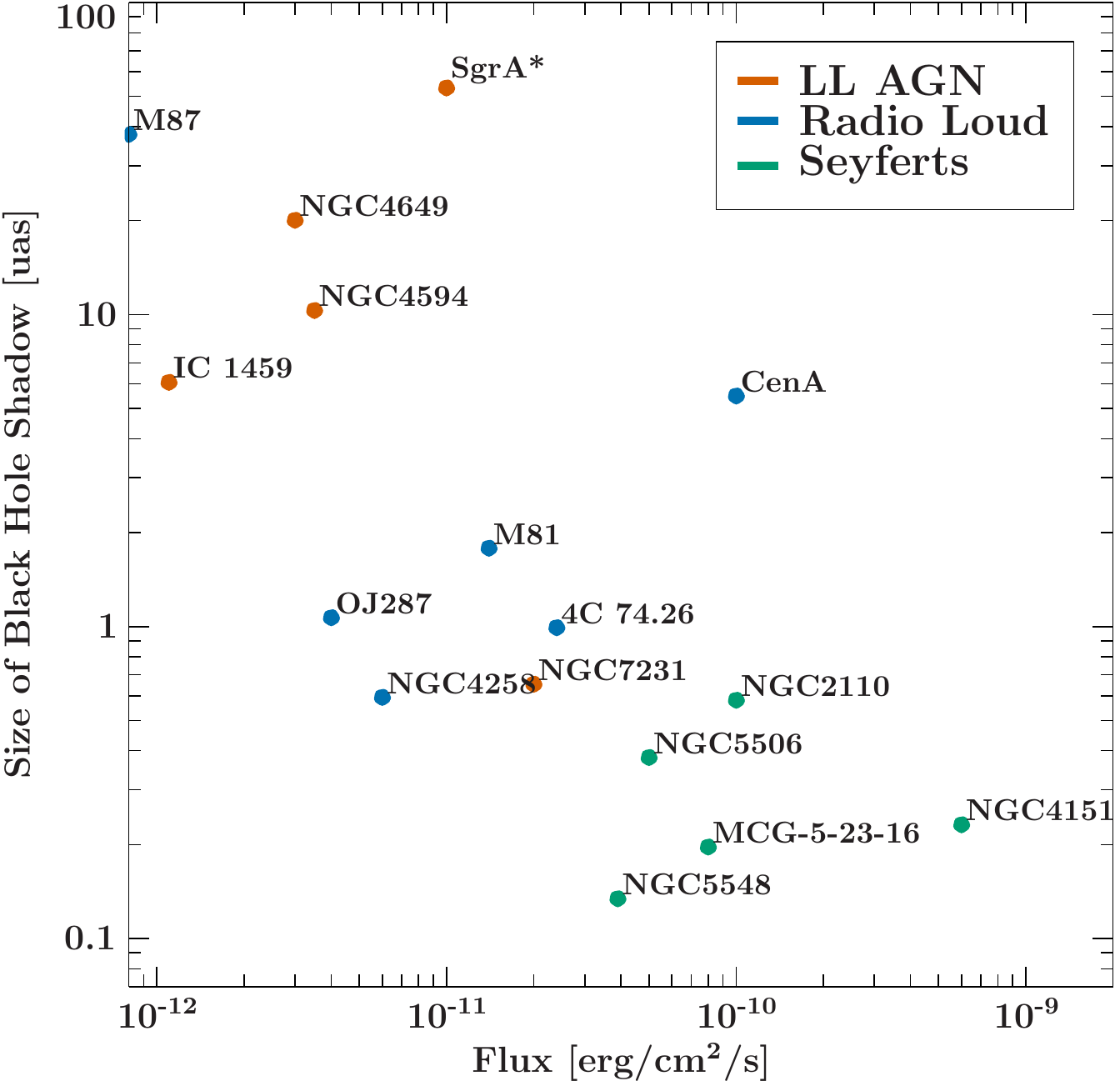}
    \caption{Left: Energy-resolved iron~K$\alpha$ image of a SMBH accretion disk that reflects X-ray photons from a corona, shown for a spatial resolution of 0.1~\uas. It shows a unique map of photon rest energy changes due to Doppler and GR effects. The precise shape of the constant energy contours of the iron K-shell emission would provide a unique test of our understanding of the dynamics of matter in strong-field gravity. The simulation assumes a BH with $M=10^9~M_\odot$, dimensionless spin $a=0.998$ at a distance of $30$ Mpc and viewed from an angle of $i=60^\circ$. Right: Possible SMBH targets for interferometric imaging, which are divided into three categories, including low-luminosity AGN (with low accretion rates), radio-loud AGN and Seyfert galaxies and potentially representing the full range of AGN accretion and outflow states. Targets are displayed depending on the estimated size of their BH shadow on the sky and X-ray flux (given in the 2-10\,keV band).}
    \label{fig:imaging-corona-energies}
\end{figure}

An unambiguous test of the theory must rely on pure geometrical properties of the curved spacetime and use matter only as a source of radiation without relying on any property of the accreting medium. Einstein himself suggested three tests to verify his theory: Mercury perihelion precession, bending of light in gravitational fields and the gravitational redshift. The latter two satisfy the above condition. Being able to observe the shape of the gravitationally bent image of a thin accretion disk and to track the change of the iron K-alpha line energy along a constant radius ring would be unique and powerful checks on the theory that are only possible using XRI (see Fig.~\ref{fig:imaging-corona-energies}).

The unique combination of interferometric imaging with the intrinsic spectral resolution of X-ray detectors will enable powerful measurements of relativistic dynamics and potential tests of general relativity, through imaging-spectroscopy of the accretion disks of supermassive BHs (SMBH) in nearby active galaxies.  X-ray imaging can clearly reveal the position of the hot X-ray emitting corona as well as how far the accretion disk extends down towards the BH. The position of the innermost stable circular orbit (ISCO) is strongly dependent on the BH rotational angular momentum (or `spin') and can reach almost to the event horizon for a maximally rotating BH or stop at about six gravitational radii\footnote{The gravitational radius for a BH of mass $M_{\rm BH}$ is $r_{\rm g}=GM_{\rm BH}/c^{2}\simeq 1.5(M_{\rm BH}/M_{\odot})$~km} for a non-rotating BH, or even further out for a counter-rotating BH.  Direct kinematics of the orbiting disk material down to these radii will be provided by the spatially resolved iron line profile: the spectrum from each resolved pixel on a disk image will show an iron line redshift that encodes the local velocity and gravitational redshifts. 

Measurement of the redshift versus position on the disk will not only enable the detailed study of the fluid dynamics of accretion flows in strong gravitational fields, but will also provide clean measurements of such strong field GR effects as light bending and gravitational redshift. X-ray spectroscopy and timing can be used to infer BH mass and spin, and even test for deviations from the Kerr metric, but this is all contingent the assumption of Keplerian orbits outside the ISCO and of free-fall inside of the ISCO, as well as degenerate to some extent with the assumed illumination pattern of the corona on the disk. Only with XRI will it be possible to test all these assumptions, break the dependence of the results on coronal illumination and therefore reach the long-held goal of using accreting BHs to explore and check some of the detailed predictions of GR for the behaviour of matter in strong-field gravity, e.g. through the radial dependence of gravitational redshift, the dependence of redshift on orbital frequency and the trajectory of null-geodesics. 

\subsection{Cosmology and cosmic evolution}
The effect of cosmic expansion on angular size means that 1~pc subtends more than $100$~\uas\ across the entire observable universe. This means that a single-spacecraft XRI has remarkable power as a probe of cosmology and cosmic evolution, for any compact sources it can detect out to large distances. For example, many luminous X-ray sources at cosmological distances, from quasars to explosive transients, show significant X-ray variations which may reverberate off gas in their near environs, producing a light-echo at a light-travel delay of a few years or less. A single-spacecraft XRI could resolve these light echoes and the flares that produced them and use the delay to give a direct estimate of the linear distance at the source which corresponds to the measured angular distance. If the light-echo geometry is relatively simple or its structure is well-mapped, such measurements would provide a powerful new trigonometric method to determine the Hubble constant $H_{0}$ over a range of redshifts.

Besides measuring light-echoes at cosmological distances, XRI could address the existence and nature of binary supermassive BHs which, due to the cosmic history of galaxy mergers, are expected in potentially large numbers in the nuclei of galaxies. If both SMBHs are accreting X-ray sources they can be resolved by an XRI, and over time their orbital motion can be measured to confirm their binary SMBH nature. A single-spacecraft XRI can potentially identify binary SMBHs with $>1$~pc separation across the entire observable universe (if they are sufficiently luminous), with the resulting demographics providing crucial information on the nature of the merger process.  For example, do many SMBH mergers stall at small radii due to lack of an effective mechanism for angular momentum loss (the so-called `final parsec problem' \cite{MilosavljevicMerritt2003})? If so, we would see many more $\sim 1$~pc binaries than if merging is efficient. It is also interesting to note that any close binary SMBH whose orbits can be resolved up to $\sim 100$ Mpc away will also radiate gravitational waves at frequencies $> 10^{-9}$~Hz, which might be detected in the next couple of decades by pulsar timing arrays (PTAs) \cite{Mingarelli2017}.

\subsection{Gamma-ray burst afterglows and gravitational wave events}
Some of the most fundamental information on the physics of gamma-ray bursts (GRBs) and related gravitational-wave events have been obtained from resolving them spatially. Since they are very compact and at high redshift, this requires the highest of angular resolutions, and thus has so far succeeded only in a few exceptionally nearby cases using radio VLBI. For these few cases, it allowed us to directly measure the expansion speed of one long GRB \cite{Tayloretal2004}, and enabled the unambiguous determination that a relativistic jet was ejected in the neutron star merger (and short-GRB) GW170817 \cite{Ghirlandaetal2019}, as well as the determination of some key physical parameters.

Resolved GRB observations are important because much of the uncertainty in our understanding of energetics, dynamics, and rates of GRBs is related to the fact that they are point sources of which we can only measure the integrated emission, and then model our way to understanding them.  If they can be resolved spatially, and as a function of time, we can directly measure the expansion rate and opening angles of their collimated explosions, and thereby directly measure their total energies. Crucially, these estimates could be done at early times, when radio data does not resolve them yet; combining these XRI results with late time radio data in those few case where we would have both, would provide Rosetta stones for GRB astrophysics.

XRI with a single satellite or with a 100-m multi-satellite baseline would bring size determinations of GRBs to the realm of the very regularly possible. GRB afterglows have typical X-ray fluxes of $10^{-11}$~\cgs\ after half a day and remain above $10^{-12}$~\cgs\ for about a week, which should allow follow-up with a reasonable satellite response and re-pointing time. After 1 day, the transverse size of a typical GRB afterglow is expected to be $\sim 0.02$~pc, corresponding to 2~\uas\ at $z=1$, 10~\uas\ at $z=0.1$, and 120~\uas\ for the nearby GW170817. A source as close as GW170817 is expected to occur only once every few years, but a few per cent of all GRBs are within z=0.1, and about 20\% are within z=1 \cite{LeandMehta2017}. Given the brightness of GRB afterglows, for a single spacecraft XRI, measuring sizes starting from 20~\uas\, i.e., 10\% of the interferometer resolution, should be quite feasible, implying that jet expansion dynamics and opening angles can be determined for up to 10 GRBs per year, and for most neutron star mergers within the horizon of current GW experiments. 
A multi-spacecraft constellation can follow in detail the evolution of the X-ray emitting surface size and shape of a GRB afterglow, measuring directly its relativistic expansion dynamics, emission charateristics, and constraining such currently poorly known parameters as the jet opening angle and true total energy.

\subsection{Coronae and jet launching from SMBH}
The previous 20 years of study of X-ray emission from accreting SMBH have suggested that the primary source of X-rays, the corona, is very compact and a large fraction of this emission is reflected and reprocessed by the innermost parts of the accretion disk.  These findings are confirmed by reverberation studies \cite{kar16} and microlensing observations \cite{Chenetal2011}, and often place the X-ray corona to within 10 gravitational radii from the BH. The evidence from the relativistic reflection spectra for strong focusing of the radiation from the corona on to the innermost parts of the accretion disk, also suggest that the corona is very compact \cite{WilkinsandFabian2011,Svobodaetal2012,Dauseretal2013}.

\begin{figure}[tb]
    \centering
    \includegraphics[width=0.48\textwidth]{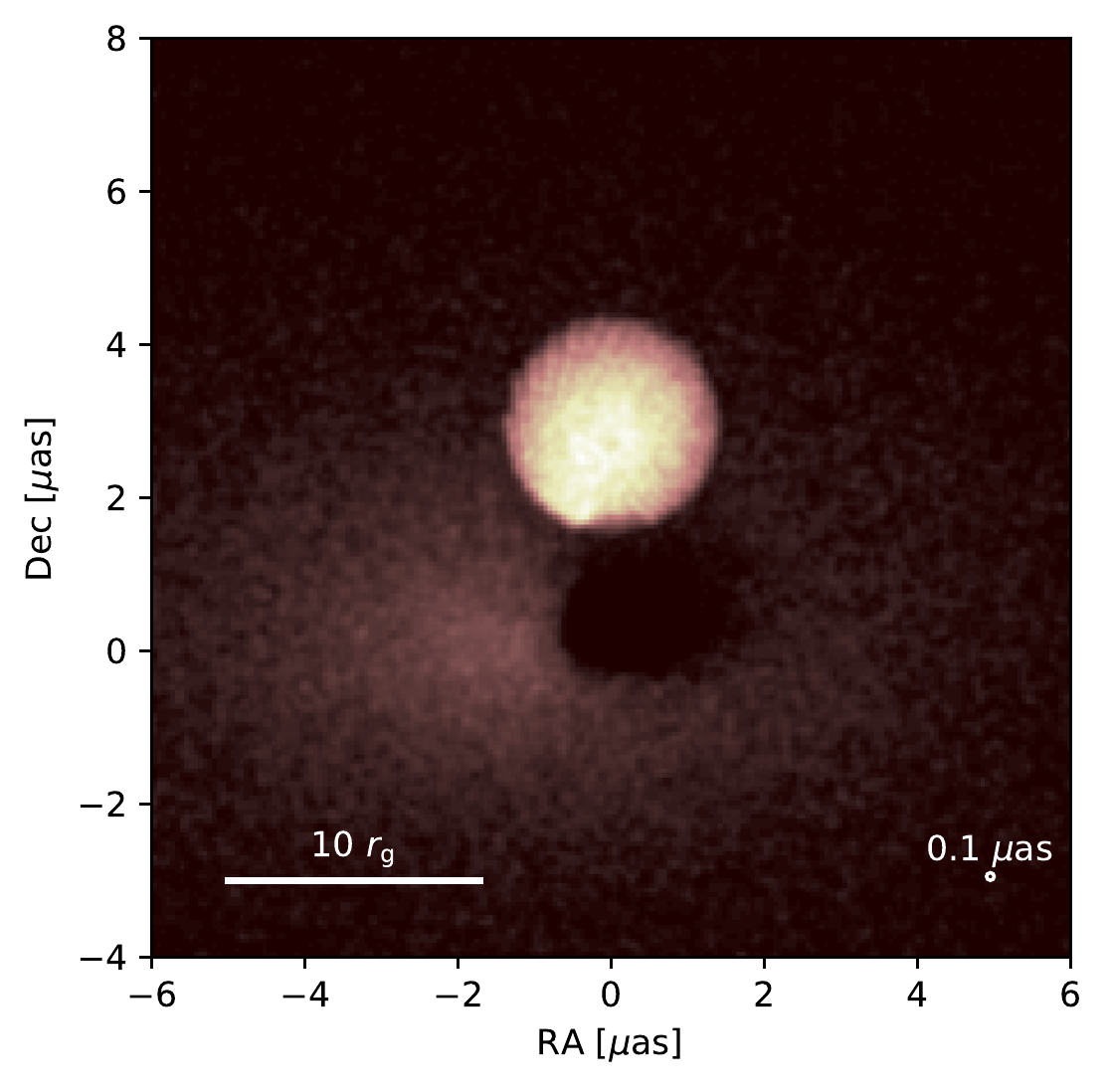}
    \hfill
    \includegraphics[width=0.48\textwidth]{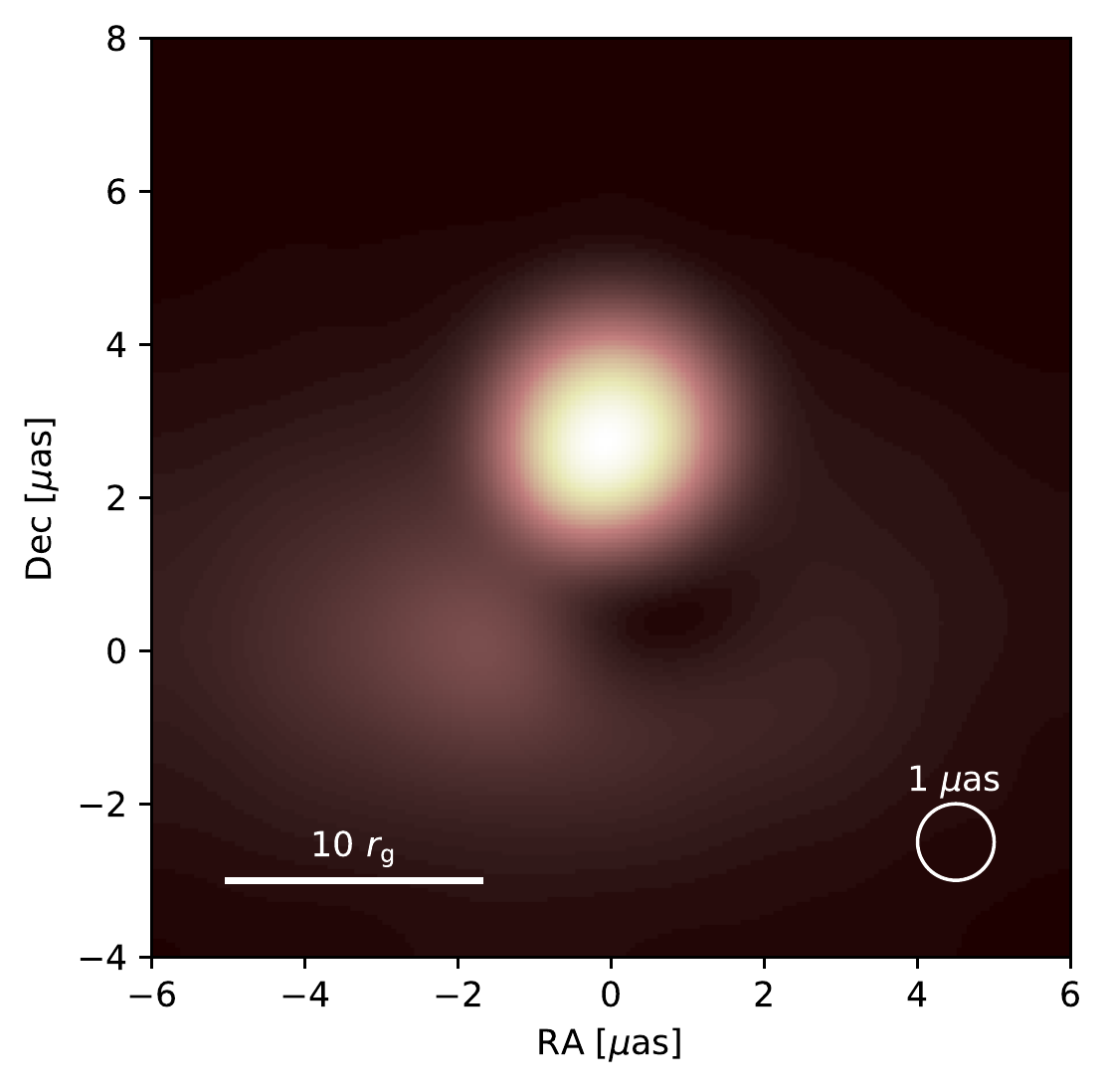}
    \caption{Image of an example geometry of an AGN core: a spatially extended hot corona is located above an accretion disk that surrounds the central BH, assuming the same system parameters as in Fig.~\ref{fig:imaging-corona-energies}. The corona is the primary source of X-ray photons that reflect from the accretion disk. \textit{Left:} Image blurred by 0.1~\uas\ FWHM 2D Gaussian (indicated by the white circle). \textit{Right:} The same model blurred by a 1~\uas\ Gaussian. Image brightness is linearly proportional to the photon count.}
    \label{fig:imaging-corona}
\end{figure}

Such compact coronae very close to the BH are however, energetically unstable \cite{fab15a} and how they are heated and store energy, perhaps through a significant magnetic field generated by (or advected through) the disk, remains an open question. Furthermore, recent reverberation studies of both an AGN and a BH X-ray binary suggest that besides a compact base, BH coronae may have a significant vertical extent, which can change over time \cite{Karaetal2013,Karaetal2019}. Therefore major questions regarding the nature, location, and origin of BH coronae remain. XRI will be able to directly measure the location and extent of the corona without ambiguity (see Fig.~\ref{fig:imaging-corona}). The direct localisation of the corona will shed light on its nature, whether it is part of a hot inner accretion flow, a multi-patch structure appearing and disappearing with magnetic reconnection in the disk, or the base of the outflowing jet. Such a discovery would enable a much deeper understanding of how accretion produces coronae and provide an essential observational check on self-consistent magnetohydrodynamic models of accretion flows in strong field gravity, which should advance significantly with computational power over the coming decades.

Relativistic jets are one of the most prominent AGN phenomena, potentially carrying equal or greater power than produced by accretion itself, but their physical origin is still highly uncertain. They are clearly detected in about 10\% of sources, so called radio-loud AGN, although they may also be present in radio-quiet AGN. A key question is whether the jet is powered by energy extraction from BH rotation (the Blandford-Znajek mechanism \cite{BlandfordZnajek1977}) or by the accretion flow (Blandford-Payne \cite{BlandfordPayne1982}). XRI will provide fundamental tests of the jet origin, by probing the basic assumption of the existence of the ISCO which underpins BH spin estimates that can be compared with jet power. Detection of an ISCO signature in a hot accretion flow would further allow measurement of BH spin for sources with powerful jets, when reflection spectral-fitting methods will not be applicable if the inner accretion flow is hot and advection dominated.  The difference between the two jet powering mechanisms could also be directly revealed in the orientation of the jet if the BH spin axis is not exactly perpendicular to the imaged accretion flow. While the Blandford-Znajek process produces a jet along the BH spin axis, the Blandford-Payne mechanism leads to jets perpendicular to the accretion flow. XRI will reveal the geometric relation between the jet, disk and X-ray corona and will allow us to literally see how relativistic jets are formed from the accretion flow.

\subsection{AGN near-environment, feeding and feedback}
The Unified Model for AGN has passed many tests in the past few years, thanks to multiwavelength imaging, spectral, and timing observations. The existence of the classical ``torus'', in the generic sense of an axisymmetric circumnuclear absorber has been confirmed, but its geometrical and physical structure has been proven to be far from being homogeneous and fully understood. There is now strong evidence for absorption/emission components on very different scales \cite{Bianchietal2012}.

{\it On scales of 100s of pc, or even kpc}, galactic dust lanes and circumnuclear tori have been imaged \cite{Malkanetal1998}, with different techniques, and determine the type 2 (in optical/UV) or absorbed (in X-rays) classification in a sizeable fraction of AGN.

{\it On the pc scale}, down to the dust sublimation radius, the standard torus postulated by AGN unification models has now been directly imaged in a few sources with IR interferometric techniques \cite{Jaffeetal2004} and its presence in nearby active galaxies is further suggested by dust reverberation mapping in the near-IR \cite{Suganumaetal2006}, and by X-ray reflection spectra.

{\it On the sub-pc scale}, dust-free gas along the line of sight has been revealed through X-ray absorption variability in many AGN \cite{Risalitietal2005}, clearly showing that part of the observed absorption is due to Broad Line Region clouds. 

XRI will finally allow us to image the inner regions in emission, in particular by mapping the Fe K$\alpha$ emission line which characterizes the reprocessing of the nuclear radiation field from Compton-thick matter. In the brightest obscured Seyfert galaxies part of the outer emission can be barely resolved by {\it Chandra}. This ensures the possibility for XRI to image it in detail, allowing us for the first time to directly observe its clumpiness and the morphological correlations with other components at other wavelengths. This could be also done for specific clumps of the outer Narrow Line Region.

The Broad Line Region (BLR) has been extensively studied in the UV, optical and also X-ray band \cite{Costantinietal2016}. It is a thick ($N_{\rm H}=10^{23}$~cm$^{-2}$) region extending on sub-pc scales around the central source. It is gravitationally bound to the central source and most likely stratified, with the inner, more ionized part emitting X-rays ($r<0.01$~pc) up to the optical emitting part ($r<1$~pc). The boundaries of the BLR are believed to be set by the sublimation radius, which limits the inner radius of the dusty molecular torus (e.g. \cite{Landtetal2014}). XRI will be able to investigate the innermost part of the BLR, with the perspective of studying its dynamics and geometry by direct imaging. The dynamics will yield accurate measures of the SMBH mass and the spectroscopic information from each location will also determine the connection of the X-ray emitting BLR with the outer UV/optical region, determining the so-far unknown structure of the AGN on sub-pc scales. In particular, it will be possible to directly measure the ionization and density radial profiles, to be compared to specific predictions of physical models (e.g. \cite{Baskinetal2014}). With an effective area of 0.1~m$^{2}$, we will be able to access the BLR of the numerous bright ($F>10^{-12}$~\cgs) AGN at redshift $z<0.01$. For brighter objects, combined imaging and reverberation mapping, via the monitoring of the emission line changes as a function of the continuum, will be possible, helping to understand the dynamics of the region by mapping it in 3-dimensions using light-travel time to constrain the line-of-sight distances to the different regions of the emitting gas seen in the 2-D image. 

X-ray emitting/absorbing outflows from the innermost regions are a common feature in AGN and testify to the continuous trade-off between accretion and ejection in an AGN. 
Outflows will eventually escape the system or become part of a larger scale bulk mass motion that will influence the host galaxy, the so called feedback, but the efficacy of this process depends on difficult-to-measure quantities such as the total kinetic power of the outflow.
Present and planned missions can determine the outflow ionization, velocity, density and launching radius, but the total kinetic power depends also on the assumed clumpiness of the outflowing gas, which can only be measured unambiguously through direct imaging of the emitting clouds. These outflows are also variable in time, revealing a complex dynamics that reflects the BH activity, the launching mechanism of the flow and the surrounding/confining medium. XRI will be able to easily image this emitting outflowing gas, for example via O{\sc VII} ($\lambda \sim 22$~\AA) imaging in the nearby universe for sources with flux $>10^{-12}$~\cgs.

The enormous power output (both radiative and kinetic) produced by AGN can only be explained by accretion of matter on to SMBHs. Nevertheless, we still do not understand the mechanism which is responsible for angular momentum transfer in the accreting matter, especially on larger scales between the extended gas which feeds the AGN, and the accretion disk. Very often accretion on large scales is via two phases, where hot X-ray emitting plasma coexists with cold dense gas visible in the optical/UV domain. Therefore, with X-ray imaging-spectroscopy of the hot plasma in the vicinity of BHs in nearby AGN, combined with that of cold gas illuminated by the central X-ray source, we can understand much better how AGN are fed on large scales via the different gas modes, and how these modes also interact with the outflows and X-ray radiation from the central AGN. These data will provide crucial inputs for understanding the interplay of feeding and feedback in AGN, to enable a complete understanding of the role of AGN in the formation and evolution of their host galaxies and clusters.

\subsection{X-ray binaries: orbits, jets and disk winds}
The masses of accreting BHs and neutron stars in X-ray binary systems (XRBs) are usually measured through velocity curves of their orbital motion, obtained through optical or IR high-resolution observations. However this method is applicable only to a very small sample of objects, because of a number of technical limitations, including the difficulties in separating the optical emission from the two bodies of the binary, and the impossibility to measure any radial velocity for very high inclination sources. XRI will solve both these issues, by providing us with direct imaging of the binary, transforming many XRBs into ``visual binaries'', enabling orbital motion and hence fundamental system parameters to be directly measured. This challenging objective becomes feasible with a single-spacecraft interferometer. Any compact object with a Roche-lobe filling companion and an orbital period longer than $\sim1$~day will have an orbital separation larger than $\sim8\times10^{6}$~km, i.e  50~\uas\ at 1 kpc. The orbital separation increases if the binary is heavier (because of a heavy BH or a massive companion) and if the orbital period is longer. For a 30~$M_{\odot}$ binary with a 10-day orbital period, the orbital separation would be ten times larger, so resolvable at 10 kpc. A resolution of 1~\uas\ would allow us to resolve a low-mass neutron star binary (total binary mass of 3 solar masses) with a 0.1-day orbital period, 10 kpc away. That is, virtually the whole X-ray binary population would be accessible. Assuming X-ray fluxes from the irradiated companion of the order of $\sim 10^{-6}-10^{-2}$ the total X-ray luminosity from the binary system \cite{Stilletal2001}, both components of the binary will be easily detectable for most X-ray binaries at their (Eddington-limited) outburst peak.

Jets are ubiquitous in the Universe, produced by accretion across a broad range of mass and energy scales, from young-stellar objects to super-massive BHs. Yet, many unknowns limit our knowledge of the physics of jets, including the mechanism that (re-)accelerates matter into and along the jet. X-ray binaries have great potential for solving this, because of the large dynamic range of luminosity and scale which they span over accessible timescales. XRI will allow us for the first time to image directly the X-ray emitting regions of the jets in these systems, studying their evolution and dependence on the accretion properties. This will be easily done for at least one source - the microquasar SS~433 - where variable X-ray emission from collimated jets has been spectrally identified \cite{Marshalletal2002,Marshalletal2013} and located at an estimated distance from the accreting BH of about $10^{7}$~km, corresponding to an angular scale of 13~\uas\ at 5~kpc. An additional X-ray jet component has been tentatively imaged with {\it Chandra} at a distance 5 orders of magnitude larger than this \cite{Migliarietal2002}, suggesting a re-acceleration mechanism at work along the jet.  Several other sources are known to go through accretion regimes similar to the one SS~433 is permanently in, with the detection of fast discrete ejections at radio wavelengths and very high measured accretion rates. Thus, there is a great discovery potential in targeting these sources with XRI angular resolution, to discover similar resolved jet components in X-rays. 

Similar to jets, wide-angle winds are a fundamental ingredient of accretion processes. While jets are known to carry energy out to kpc scales, the mass is removed from XRBs via X-ray winds \cite{Pontietal2012}. These winds are observed in absorption in every XRB that is known to be at high inclination and with a relatively long orbital period ($>20$~hours) indicating an equatorial geometry and the need for a large disc. They have the potential to significantly destabilize the accretion discs that launch them \cite{Munozdariasetal2016,Shieldsetal1986}.
The launching radius of the winds is a fundamental parameter to determine the launching mechanism. Winds are estimated to be launched at or close to the Compton radius. For the bright BH V404~Cyg, \cite{Kingetal2015} estimate a launching radius in the  $(4-30) \times 10^{6}$~km range, implying that with 10~\uas\ resolution we would be able to explore the upper range for most of the known XRBs, while we would need 1~\uas\ resolution to explore the full range, yielding unprecedented constraints on theoretical wind models. Furthermore, by imaging the winds and detecting them in emission, we could also study them in low-inclination systems and not only those with a favourable line-of-sight. With a higher resolution than 1~\uas\ we would be able to identify the position of cooler and warmer components of the wind, constraining the geometry of the wind and disk atmospheres also for smaller systems.

\subsection{Massive stellar winds}
Stellar winds are key ingredients for understanding massive stars and massive star binaries: they can reach mass loss rates up to 10$^{-5}\,M_\odot$\,yr$^{-1}$ and velocities up to 3000\,km/s. They define the evolutionary pathways of massive stars (the progenitors of explosive transients and gravitational wave sources), trigger and inhibit star formation in solar nurseries and enrich their environment \cite{Kudritzki_2000a}. The structure of these winds is, however, poorly understood: line-driving is prone to instabilities, leading to a complex, inhomogeneous wind structure \cite{Martinez-Nunez_2017a}. X-ray emission has been observed from individual stars but especially also from interacting wind binaries \cite{Rauw_2016a}.

If a massive star is in a binary with a BH or neutron star (i.e., in a high mass X-ray binary, HMXB), the stellar wind drives changes in the accretion rate and thus the system's X-ray emission. The interaction of this emission with the wind material can be used to study the wind itself: its geometry, porosity, mass-loss rate and interaction with the compact object, e.g., accretion and photoionization wake \cite{Watanabe_2006a,Grinberg_2017a,Martinez-Nunez_2017a}. Understanding wind accretion in HMXBs is also paramount as we expect these systems to be a necessary step towards double degenerate systems whose mergers we observe as gravitational wave events \cite{Giacobbo_2018a}. Our current understanding of the geometry of interacting binary systems and HMXBs relies on toy models that are not sufficient to explain the complexity of these systems \cite{Watanabe_2006a,Martinez-Nunez_2017a}. Constraints on wind models and answers to longstanding questions, such as whether the X-ray irradiation of the wind in HMXBs suppresses wind-launching and how far it changes the wind structure \cite{Krticka_2018a}, are only possible if there is a leap forward in our understanding of the geometry of these systems.  

A resolution of 100~\uas\, achievable with a single spacecraft interferometer, corresponds to $\sim$45\,$R_\odot$, i.e., the typical size of an O/B-supergiant, at a distance of $\sim$2\,kpc, the distance of some of the key HXMBs Cyg~X-1 and Vela~X-1 and the interacting binary $\eta$~Car. It would allow us to resolve the largest scale of the wind structure in these systems. In Vela X-1, where high and low ionization lines from different elements \cite{Watanabe_2006a,Grinberg_2017a} are simultaneously present, we will thus, for the first time, be able to trace whether the cold material is concentrated in the accretion wake or directly embedded in the wind as denser, colder clumps. In interacting binary systems, we will be able to directly image the X-ray emitting interaction region, similarly to the optical \cite{Weigelt_2016a}, constraining the location and structure of the most energetic interactions. Tracing the variability of the radiation zone through multiple observations in both kinds of systems will give us further constraints on wind variability and thus wind driving as well as system geometry. 

\subsection{Stellar coronae and exoplanets}
Stars emit X-rays from their hot outermost atmosphere layer, the corona. All stars on the cool half of the main sequence are X-ray emitters; their coronae are heated by a mechanism that is not fully understood yet and likely includes both acoustic waves and magnetic flares, which brings the coronal plasma up to temperatures of several million~K. The coronae cool through emission in metallic spectral lines at typical X-ray energies between 0.1-5 keV \cite{Guedel2004}.

Currently the only star for which we have a spatially resolved image of its corona is the Sun. For other stars, we are limited to indirect reconstruction methods such as light curve inversion for fast rotating stars or eclipsing binaries (see for example \cite{SchmittFavata1999}). Such methods have many ambiguities and have not been able to answer many fundamental questions, such as what what the typical extent of stellar coronae is, if they are similarly structured to the solar corona, or if they can have fundamentally different surface patterns such as polar active regions.

\begin{figure}
    \centering
    \includegraphics[scale=0.08]{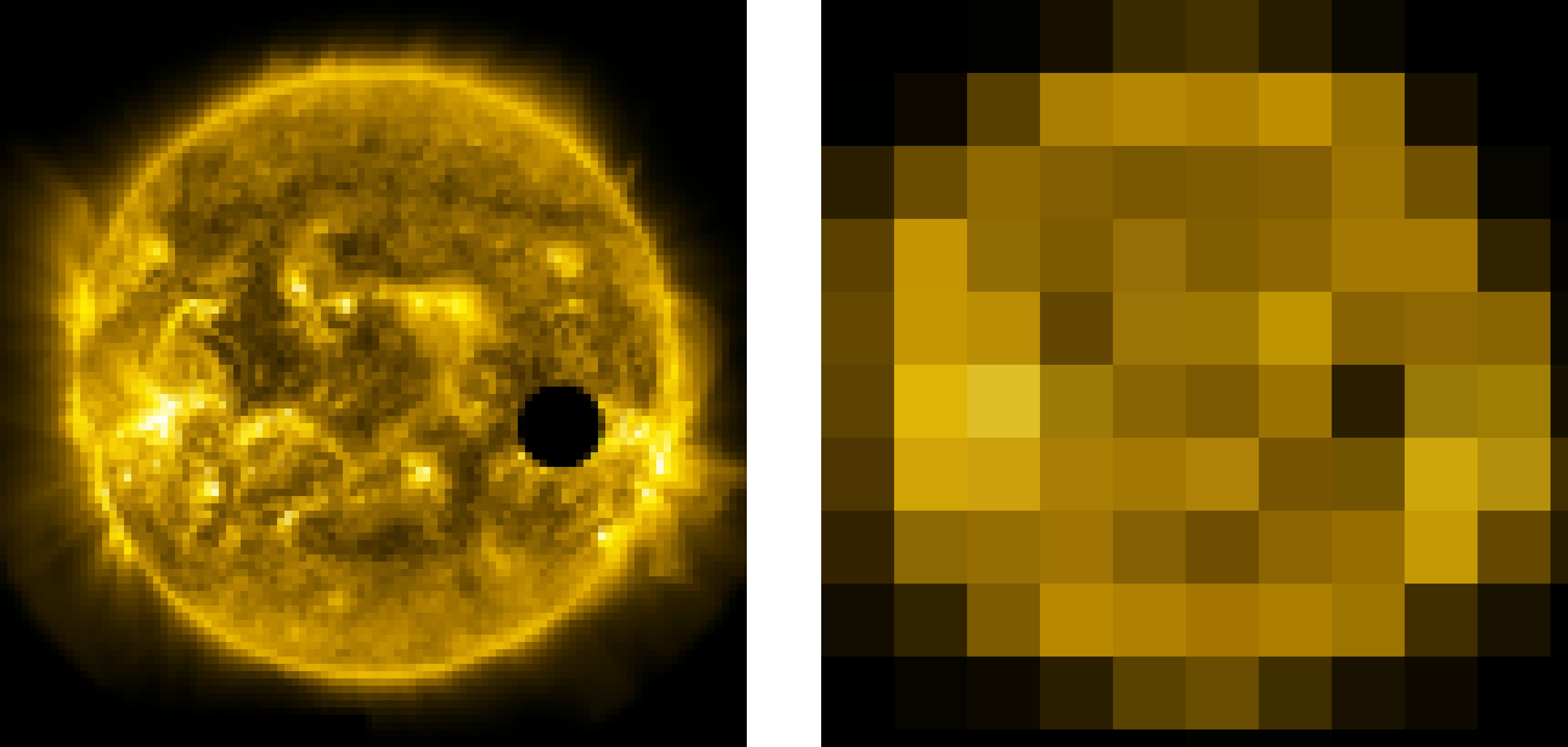}
    \caption{Simulated image of the corona of a low-mass star with a transiting exoplanet at 10 pc from the Sun, with a resolution of 10~\uas\ (left) and 100~\uas\ (right). This example uses the active solar corona (SDO AIA 171 band) as input, and assumes system parameters of the star and planet similar to the AU Mic system. AU Mic has two known transiting exoplanets and an X-ray flux of $3\times 10^{-11}$ \cgs.}
    \label{fig:stellarcorona}
\end{figure}

With X-ray observations with resolutions of 10-100~\uas, we can breach these limitations for the first time. Sun-like stars at a distance of 10~pc have an angular diameter of about 900~\uas, and lower-mass stars like M dwarfs (which often tend to be active and X-ray bright) have angular diameters of about 450~\uas\ at that distance. Within the solar neighborhood there are a good number of stars with high X-ray fluxes around $10^{-11}$~\cgs, that will lend themselves to coronal imaging at stunning resolution. Figure~\ref{fig:stellarcorona} shows how the corona of on active low-mass star similar to AU~Mic would be imaged at 100~\uas\ and 10~\uas\ resolution. At 100~\uas\ a basic analysis of coronal bright versus dark regions would be possible, and at 10~\uas\ coronal structures can be studied in detail.

A particularly exciting prospect is the possibility to study planets around stars other than the Sun (exoplanets). Statistically, cool stars have on average at least one exoplanet \cite{DressingCharbonneau2015}, and some fraction of stars hosts planets in extremely close orbits of only a few days duration \cite{Howardetal2010}. Planets that cross the stellar disk in our line of sight are called transiting planets and provide the opportunity to study the extent and composition of their atmospheres as they are back-illuminated by the star. A current limitation of the field is that, as long as stars have to be treated as point sources, any surface features of the star manifest themselves as systematic errors in the measured exoplanet parameters \cite{Czeslaetal2009}.

With a resolved stellar atmosphere, the planetary transit can be imaged across the stellar disk. This is both extremely valuable as an X-ray experiment and for exoplanet science overall. On the one hand, the planet’s X-ray transit depth measures the extent and density of the outer planetary atmosphere \cite{Poppenhaegeretal2013}, and with spatially resolved observations this can be disentangled from localized stellar brightness changes in the corona. On the other hand, simultaneous multiwavelength observations could also use resolved-disk X-ray transits to detrend observations at other wavelengths through reconstructions of the stellar chromosphere and photosphere based on the coronal structure. As uncertainties in stellar parameters are one of the main limiting factors in studying exoplanet atmospheres today, this would constitute a major breakthrough in our understanding of exoplanets.

\section{Instrument design, technological progress and roadmap}
\label{sec:instrument}

\subsection{Introduction}

A variety of technical items need to be realised for a successful X-ray interferometer in the future. Some of these will be addressed as part of continuing developments in other fields, but for a subset we need dedicated efforts. To draw the technological roadmap, we describe the current status of XRI and the goal we want to reach, and the technological hurdles in between. In fact, under laboratory circumstances, XRI was already demonstrated 2 decades ago, in a ground-breaking experiment by Webster Cash at the University of Colorado~\cite{Cashetal2000}. X-ray fringes were recorded for monochromatic X-ray light at 1.2 keV (1 nm) in a four mirror set-up with an effective baseline of 1 mm. 
Our ultimate goal is to achieve sub-\uas\ resolution, to unlock the full potential of XRI, e.g. to enable detailed Fe K imaging spectroscopy of the close vicinity of the event horizons of SMBH. At 6 keV this requires a baseline of more than 20 m, which is minute compared to the Earth-scale baselines required for the Event Horizon Telescope (EHT) observation of the SMBH in M87~\cite{EHTm87L2_2019}. However, bringing even this small baseline into space, with stable path-length equalization at nm-level, while scaling up the collecting area from a few square mm under monochromatic laboratory conditions to the square meter scale at multi- X-ray wavelengths for astronomical purposes, invites significant steps forward in space technology.  Therefore, a lower-resolution, single-spacecraft design is most likely to be realised within the Voyage 2050 timeframe. 
 
\subsection{Interferometry approach}
XRI, like optical interferometry, requires the direct combination of photon beams. Given the very short X-ray wavelengths and moderate photon count rates, the method of Intensity interferometry \cite{Hanburyetal1954} probably cannot be used, so we will focus on phase interferometry, using an approach based on the well-known Michelson interferometer.

In each X-ray Michelson interferometer the path lengths of the interfering beams must be equalized to a fraction of a wavelength ($\sim \lambda/20$) to have good fringe visibilities, which, for 6 keV light, requires equalization to 10 pm. This is similar to the resolution to be obtained in ESA's LISA mission~\cite{Audley2010} for paths of a million km, hence not unfeasible. Two choices are available for the main optical elements: 
\begin{itemize}
\item \textit{Mirrors:} X-ray mirrors require grazing incidence for effective reflection, where incidence angles $\theta_{\it g}$ of typically less than a degree are needed, depending on photon energy and mirror material. Various studies have been performed, ranging from a single spacecraft approach~\cite{Willingale2004} with a relatively modest resolution of $\simeq 100$ \uas, a basic two spacecraft configuration~\cite{Cashetal2004} at a separation of a few 100 km, to a full constellation of collecting spacecraft plus a detector spacecraft \cite{Maxim2004}, extending over 20,000 km and providing an angular resolution of 0.1 $\mu$as. The latter separations sound extreme but might be significantly mitigated by the optics design (see below).
\item \textit{Gratings:} XRI with gratings employs the Talbot effect~\cite{Talbot1836} to make a self image of the grating onto the detector plane. A recent study~\cite{Hayashidaetal2018} however indicates that this method gives only very limited angular resolutions of the order of 100s of mas, and is for the moment not further considered. 
\end{itemize}

\subsection{Areas for technical progress}

\subsubsection{Optics design}

\begin{figure}
    \centering
    \includegraphics[scale=0.5]{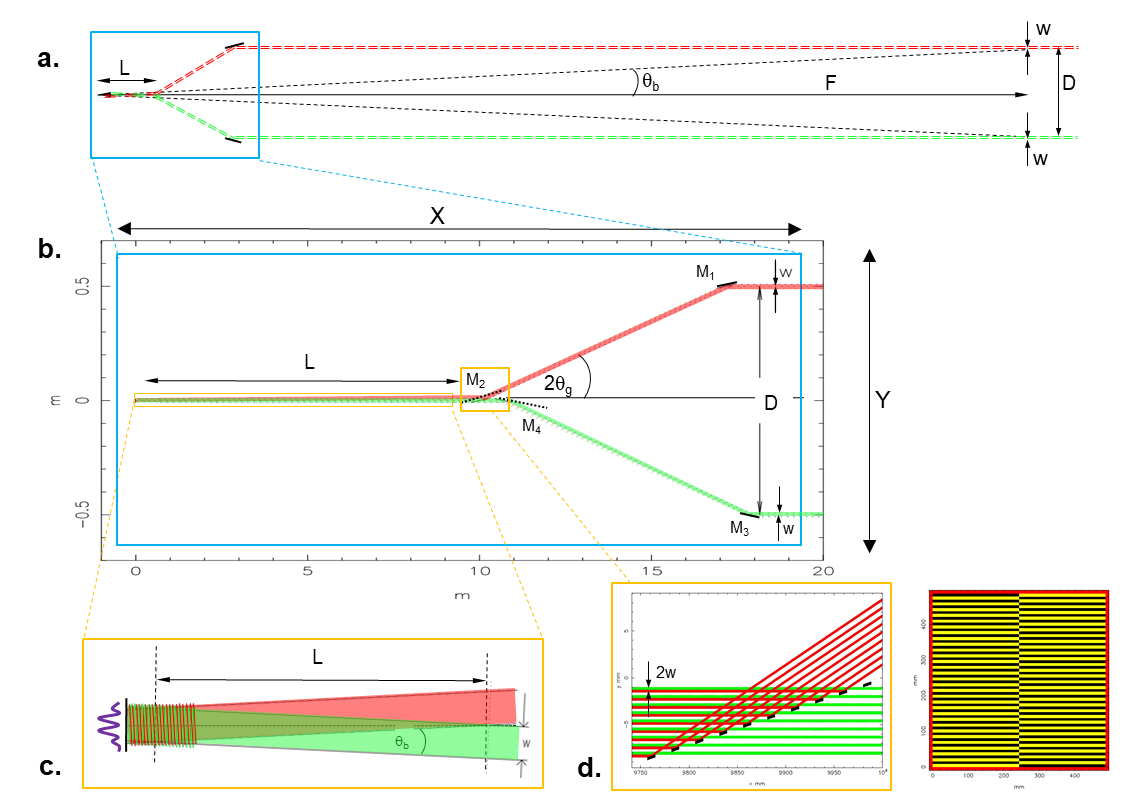}
    \caption{Schematic representation of the optics of the X-ray interferometer, at various levels of detail. 
             a.~Overview, comparing the classical and the telephoto geometry. Here $D$ is the effective baseline, $F$ the focal length, $w$ the width of a single beam and $L$ the length over which the interference of X-ray beams takes place. b.~Zoom in on the telephoto geometry, showing how flat mirrors M1, M3 and M4 and slatted mirror/beam-combiner M2, plus the detector plane fit inside a compact volume, defined by dimensions $X$ and $Y$, which is roughly compatible with the dimensions of a launcher fairing. c.~Zoom in on the interference of one pair of beams, explaining the relation $\theta_{\rm b} = w/L$. d.~Side view and front view on the slatted mirror/beam-combiner M2.}
    \label{fig:diagrams}
\end{figure}

The maximum achievable angular resolution of an interferometer is given by $\Delta\theta \simeq \lambda/2D$, where $D$ is the baseline between the telescopes collecting the two photon beams. Given the dimensions of, say, an Ariane 6 fairing, and the challenges posed by deployable and large rigid-body systems for interferometry~\cite{Dennehyetal2018}, we consider that a single spacecraft option can support 1~m baselines and is limited to an angular resolution of 100~\uas\ for 10 \AA\ light (1.2 keV) and 20~\uas\ for 2 \AA\ (6 keV) light.
Fig.~\ref{fig:diagrams} shows a specific implementation \cite{Willingale2004} of the X-ray Michelson interferometer based on flat mirrors, which contains also all essential ingredients of an X-ray phase interferometer. The geometry is governed by two angles, $\theta_{\it b}$ the angle between two interfering beams (see Figure \ref{fig:diagrams}c), and $\theta_{\it g}$. The telephoto design makes use of the ratio between these two angles to achieve a considerable compactification of the optical configuration compared to that of earlier XRI concepts: $F \simeq$ 40 km whereas $X \simeq$ 20 m. It is not possible to make $\theta_{\it b}$ arbitrarily large, as the fringe spacing on the detector plane, ${\Delta}y_{\it f} = \lambda / \theta_{\it b}$ scales inversely with it \cite{Willingale2004}, so that it is limited by detector pixel size for effective fringe-sampling (see below). The use of flat mirrors implies that the width of the beam at the entrance is equal to the beam width on the detector. Focusing optics are not considered here as current best technologies (Wolter-type optics on {\it Chandra} and {\it Athena}) are 4 - 5 orders of magnitude removed in surface accuracy from diffraction limited imaging, which is equivalent to the condition for phase interferometry. Hence, in order to achieve sufficient collecting area to reach sensitivities interesting for astronomy, massive parallelization of interfering beams is needed in combination with large area detectors. 

The design we show in Fig.~\ref{fig:diagrams} could be adapted to larger baselines by placing the mirrors M1 and M3 on board separate `collector' spacecraft, with many baselines formed by a constellation of those spacecraft plus the detector spacecraft (containing the corresponding M4, M2 mirrors and the detector). Such a configuration could achieve sub-\uas\ resolution without the very large spacecraft separations of the earlier XRI constellation designs. 

The required optical quality, in particular mirror roughness, for 1.2 keV (= 1 nm) light would be $\lambda$/200 rms (at $\lambda$ = 633nm), and for 6 keV (= 0.2 nm): $\lambda$/1000 rms. Although challenging, these requirements for flatness are not excessive.
A key optical element of the design in Figure~\ref{fig:diagrams}b is the slatted mirror concept \cite{Willingale2004}, on which mirror M2 is based, and which are schematically explained in Figure~\ref{fig:diagrams}d. It can be manufactured with lithographic techniques also applied for Athena optics, and facilitates the massive parallellization of narrow interfering beams called for by XRI, due to the short coherence lengths at X-ray wavelengths. A demonstrator prototype of a slatted mirror has already been produced and tested at optical wavelengths with promising results~\cite{Willingaleetal2005}.

\subsubsection{Imaging sensitivity}
Due to the shallow grazing incidence angles, each system of mirrors (and corresponding detector) which make up each baseline present relatively low collecting area (perhaps 50~cm$^{2}$, \cite{Willingale2004}) which is further reduced by a factor of $\sim 2$ when accounting for the mirror X-ray reflectivity and detector efficiency. However, the small footprints of the mirror/detector systems should allow the single-spacecraft/detector-spacecraft volume to be tiled with many such systems to provide many baselines, efficiently constructing images by filling the Fourier (so-called $uv$) plane with different spatial frequencies and directions. Provided that the pointing direction can be kept stable, rotation of the spacecraft about the optical axis could further fill the $uv$ plane (akin to the use of Earth's rotation in radio aperture synthesis).

An idealised X-ray interferometer is limited in sensitivity by photon counting statistics, so that (as for optical/IR interferometers in space), the imaging signal-to-noise ratio ($SNR$) is close to uniform across the image, with the rms noise per resolution element $\sigma_{\rm rms} = \sqrt{C_{\rm imaged}/2}$~counts, where $C_{\rm imaged}$ is the total number of detected X-ray photons from the imaged region \cite{PrasadKulkarni1989}.  Assuming at least {\it XMM-Newton} EPIC-pn-like collecting area, which should be easily realised on a single spacecraft, a `typical' AGN-type source with power-law spectrum of photon index $\Gamma=2$, Galactic column density $2\times 10^{20}$~cm$^{2}$ should provide a net count rate of $(F_{2-10 keV}/10^{-12})$~count/s over the full 0.1-10~keV energy range, where $F_{2-10 keV}$ is the flux in the 2-10 keV band in \cgs. 

In principle then, single sources as faint as $10^{-15}$~\cgs\ can be detected and imaged within a day, however the sparse aperture coverage demands special care with stray (that is, not interferometrically imaged) X-ray light, which potentially sets the limiting sensitivity of the instrument.  The non-imaged stray light background scales with the X-ray flux outside the imaged field of view which reaches the detector of area $A_{\rm det}$, which for a uniform X-ray sky background scales with an etendue $A_{\rm det}\Omega_{\rm stray}$, where $\Omega_{\rm stray}$ is the solid angle of the sky exposed to the detectors (and may be much larger than the imaged area of sky). Thus, increasing detector area does not automatically increase the sensitivity if it is limited by the stray light and some form of collimator may be required, provided that it does not substantially diffract the interfering beams. The effect of stray light will be similar to that of dark current in IR interferometry \cite{Haniff2007}, reducing the $S/N$ by a factor $\sqrt{C_{\rm stray}/C_{\rm imaged}}$, where $C_{\rm stray}$ is the total counts collected as stray light. Given the typical X-ray sky background and geometry of the single-spacecraft design, we do not expect the stray light flux to exceed $10^{-11}$~\cgs, even in the absence of an advanced collimator design, so that a large (factor $>10$) reduction in sensitivity will probably only be seen for targets with flux $<10^{-13}$~\cgs.

\subsubsection{Detector and Field of View}
Effective fringe sampling at the detector requires detector pixel sizes of 30~$\mu$m or better, which are readily achievable currently. However, the use of flat mirrors implies that the entrance aperture is mapped on to the detector without a magnification factor. The achievable detector area could thus eventually become a driver for the effective collecting area. The visual CCD array system on Euclid, however, demonstrates that large area detectors of (in this case 0.58 $10^9$ $\times$ (12 $\mu$m)$^2$ =) 800 cm$^2$ are already available~\cite{Cropperetal2018}. Assuming the 30 $\mu$m pixels in~\cite{Willingale2004} instead of the 12 $\mu$m for Euclid would bring us already to 0.5 m$^2$. 
In order to resolve fringes, the detector needs to have some intrinsic energy resolution $\Delta E$, where the number of discernable fringes $N_f \simeq E/\Delta E$. X-ray Si-based pixels offer a resolution in the range of 110 eV (around 1.1 keV) to 150 eV (around 6 keV)~\cite{Struederetal2001}, implying $N_f$ in the range from 10 to 40.
This resolution is sufficient to construct a useful X-ray interferometer. However, there are good reasons to seek a higher energy resolution:
\begin{itemize}
    \item \textit{Astrophysics:} For many astrophysical applications it will be important to measure velocities of resolved regions of plasma to significantly better than $\sim 1000$~km~s$^{-1}$.  Also, spectrally resolved K, L and M lines provide the fingerprints for the elementary composition of the plasma, while He-like triplets offer a wealth of information about the physical circumstances in the plasma where the light is emitted (temperature, density). 
    \item \textit{Field of view:} With the number of fringes also scales the field of view: FoV $\simeq$ $N_f$~$\Delta \theta$. A larger ratio of FoV to resolution greatly helps the localisation of the source as well as enabling images of larger fields.
\end{itemize}
The current best technology is based on Transition Edge Sensors (TES), which offer an energy resolution of 2 - 2.5 eV across a wide range of energies, implying an $R > 2500$ at 6 keV. TESs operate at a temperature of $\simeq$100 mK, and thus require a cryogenic infrastructure. The most ambitious TES instrument proposed so far, for the proposed {\it Lynx} X-ray observatory, consists of O(10$^5$) 50 $\mu$m pixels, adding up to only 2.1 cm$^2$~\cite{Bandleretal2019}. This thus invites a breakthrough technology in either X-ray detector read-out, focused optics for XRI, or massive parallellisation of cryogenic systems. It is worth stressing however that fringe sampling is only required in one dimension, along the beam baseline, so novel forms of detector could be developed, with many times fewer pixels, but where the pixels are highly elongated perpendicular to the sampling direction. Such novel designs of high-resolution 1-D detectors may provide fewer challenges than with conventional 2-D concepts.

\subsubsection{Stability and station keeping} 
Thermal mechanical stability is crucial, even for a single-spacecraft option to achieve a level of, e.g. 30 $\mu$as, both in beam positioning (on time-scales longer than an observation) and pointing stability (on time-scales shorter than an observation). Without advanced compensation, materials with coefficients of thermal expansion (CTEs) of  $10^{-7}$/degree need to be kept isothermal at the level of $10^{-3}$ degrees for structures of order 1 meter across~\cite{Gendreauetal2004}. Micro-vibrations affect the stability, of pointing and path length, as a function of time. On SIM, the Stellar Interferometer under study at NASA until 2007, the requirement on relative attitude knowledge obtained by each interferometer over all time scales up to 1 hour was 0.2 mas RMS~\cite{Dennehyetal2018}. This level was required for SIM's \uas\ precision absolute astrometry and was to be achieved by a multi-disciplinary design approach including a network of metrology systems and its feasibility was verified with a testbed. For a single spacecraft XRI, at least an order of magnitude improvement is required.

By arranging the mirrors in a periscope configuration, the requirements on the accuracy and stability of location and orientation of the mirrors is relaxed by several orders of magnitude along 2 of the 3 axes~\cite{Shipleyetal2003}. This comes at the expense of two additional reflections per beam (possibly only one for the telephoto geometry), but is probably indispensible when the interferometer is distributed over multiple spacecraft. These extra reflections, at an 80$\%$ throughput each, are comparable to the 50$\%$ throughput of the slatted mirror/beam-combiner M2 in Figure~\ref{fig:diagrams}b.  

The required absolute positioning accuracy of the optical elements depends on $\lambda$, but their realization depends on system trade-offs which can be made in future studies.

\subsubsection{Localisation of sources and pointing}
Apart from the internal stability of the optical configuration, bringing the target inside the field of view and keeping it there asks for advanced pointing technology. Cash \cite{Cash2003} and Gendreau \cite{Gendreauetal2004} have argued that star tracker technology, either in the visible or X-ray, is unlikely to reach the 30 $\mu$as level, if only due to stellar proper motion, but advanced gyroscope technology, such as e.g.~already applied on Gravity~Probe~B, will reach the required accuracy.
XRI will also come with additional questions for astrometry: homing in on a source whose X-ray position is known to 0.5 arcsec using a field of view of 1 mas may require a lot of scanning time prior to the actual observation, so it will be valuable if target positions are already known to the sub-mas level via other wavebands. For stars and the multi-wavelength spectra of AGN and many other X-ray sources, this should prove straightforward, but for sources only detected in X-ray or other high-energy emission it will be challenging, perhaps necessitating additional lower-resolution X-ray interferometers to guide the highest resolution instrument.

\subsection{A possible roadmap}
\begin{figure}[tb]
    \centering
    \includegraphics[width=1.0\textwidth]{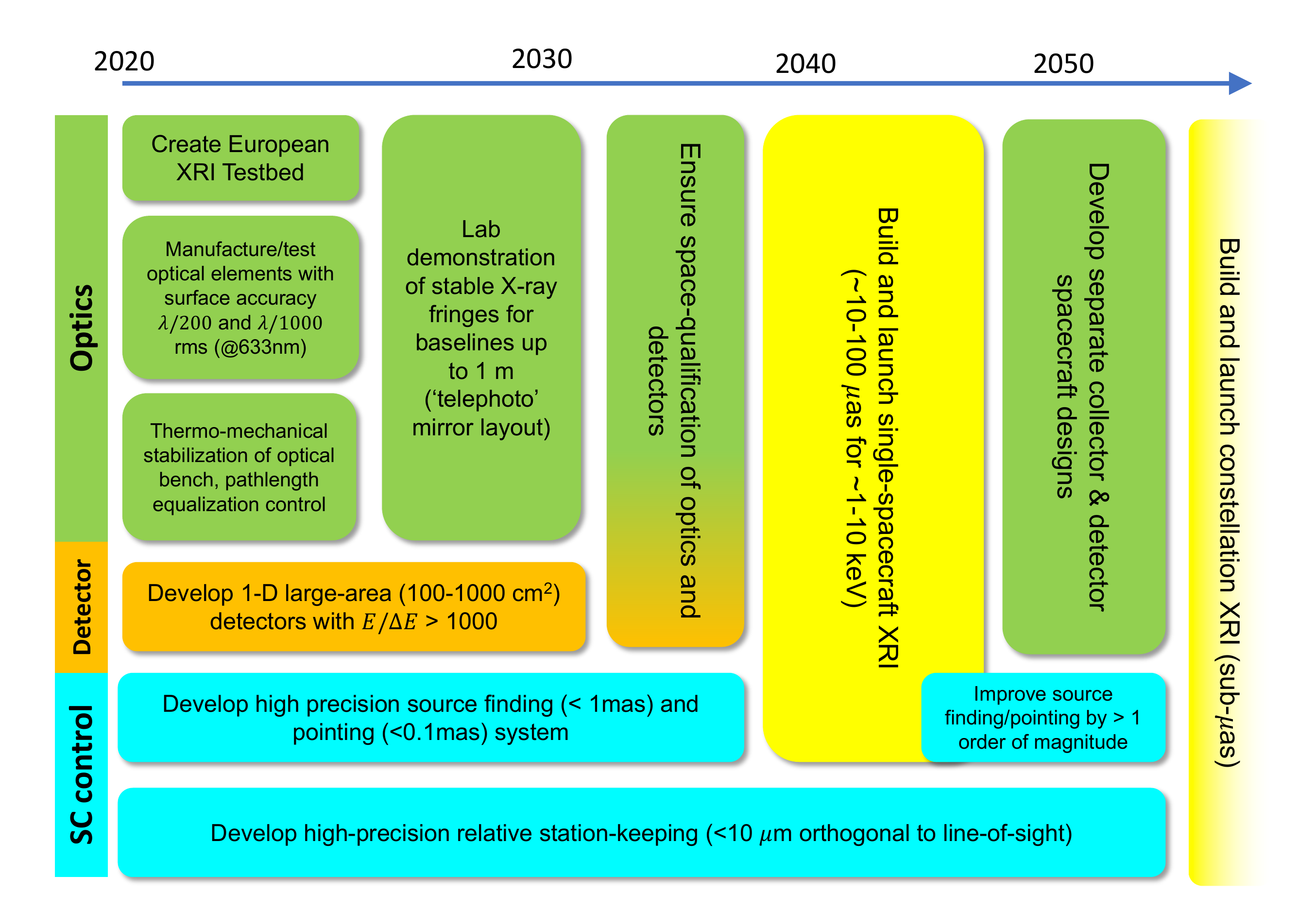}
    
    \caption{Possible roadmap showing key developments towards sub-\uas\ XRI.  The time scale shown is indicative and not linear.}
    \label{fig:roadmap}
\end{figure}

Finally, we present in Fig.~\ref{fig:roadmap} a suggested plan for the sequence of (parallel) developments that must take place to realise XRI, first on a single spacecraft and eventually (with sufficient advances in spacecraft control) a constellation XRI, which can obtain sub-\uas\ resolutions. We consider that a single-spacecraft XRI may be realised within the Voyage 2050 timeframe, enabling a host of exciting science, some of which we have discussed here.  The single-spacecraft XRI would be naturally followed by a constellation at a later stage, but we stress that if sufficient progress is made on the spacecraft control aspects, there is also no reason that a constellation could not be built sooner instead of the single spacecraft XRI, to allow a fast track to our ambition of imaging spectroscopy down to a supermassive BH event horizon. It is also important to note that once a constellation XRI is realised, further increases in resolution can likely be achieved by merely incremental improvements in spacecraft control, so that the detector baselines may be further increased, e.g. to reach the sub-nanoarcsecond resolutions that can resolve stellar mass BH event horizons and even neutron star surfaces. At that stage, X-ray astronomy will have truly shaken off the limitations of spatial resolution.

\newpage

\bibliography{xri_wp.bbl}

\newpage

\noindent {\Large \textbf{Proposing team}} \\
\\
\noindent Phil Uttley\footnote{p.uttley@uva.nl} (University of Amsterdam, Netherlands) \\
\bigskip
Roland den Hartog (SRON, Netherlands) \\
\noindent Cosimo Bambi (Fudan University, China) \\
Didier Barret (Universit\'{e} de Toulouse, CNRS, IRAP, France) \\
Stefano Bianchi (Universit\`{a} degli studi Roma Tre, Italy) \\
Michal Bursa (Astronomical Inst. of the Czech Academy of Sciences, Czech Republic)\\
Massimo Cappi (INAF, OAS Bologna, Italy) \\
Piergiorgio Casella (INAF, OA Roma, Italy) \\
Webster Cash (University of Colorado, USA) \\
Elisa Costantini (SRON, Netherlands) \\
Thomas Dauser (Remeis Observatory \& ECAP, Univ. Erlangen-N\"{u}rnberg, Germany) \\
Maria Diaz Trigo (ESO, Germany) \\
Keith Gendreau (NASA/GSFC, USA) \\
Victoria Grinberg (Eberhard Karls Universit\"{a}t T\"{u}bingen, Germany)\\
Jan-Willem den Herder (SRON, Netherlands) \\
Adam Ingram (University of Oxford, UK) \\
Erin Kara (MIT, USA) \\
Sera Markoff (University of Amsterdam, Netherlands) \\
Beatriz Mingo (The Open University, UK) \\
Francesca Panessa (INAF, IAPS, Italy) \\
Katja Poppenh\"{a}ger (Leibniz Institute for Astrophysics Potsdam, Germany) \\
Agata R\'{o}\.{z}a\'{n}ska (N. Copernicus Astronomical Centre Polish Academy of Sciences, Poland) \\
Jiri Svoboda (Astronomical Inst. of the Czech Academy of Sciences, Czech Republic) \\
Ralph Wijers (University of Amsterdam, Netherlands) \\
Richard Willingale (University of Leicester, UK) \\
J\"{o}rn Wilms (Remeis Observatory \& ECAP, Univ. Erlangen-N\"{u}rnberg, Germany) \\
Michael Wise (SRON, Netherlands)  \\

\end{document}